\documentclass[conference]{IEEEtran}
\IEEEoverridecommandlockouts
\usepackage{cite}
\usepackage{amsmath,amssymb,amsfonts}
\usepackage{graphicx}
\usepackage{textcomp}
\usepackage{xcolor}
\usepackage{algorithm}
\usepackage{multirow}
\usepackage{algpseudocode}
\usepackage{url}
\def\BibTeX{{\rm B\kern-.05em{\sc i\kern-.025em b}\kern-.08em
    T\kern-.1667em\lower.7ex\hbox{E}\kern-.125emX}}
\begin{document}

\DeclareRobustCommand*{\IEEEauthorrefmark}[1]{%
    \raisebox{0pt}[0pt][0pt]{\textsuperscript{\footnotesize\ensuremath{#1}}}}

\newcommand{\todo}[1]{\textbf{\textcolor{red}{@todo: #1}}}

\title{Advancing EEG/MEG Source Imaging with Geometric-Informed Basis Functions\\

\thanks{This work was supported by the National Natural Science Foundation of China (62001205), National Key R\&D Program of China (2021YFF1200804), Shenzhen Science and Technology Innovation Committee (2022410129, KCXFZ2020122117340001).}

}
\author{
\IEEEauthorblockN{
Song Wang\IEEEauthorrefmark{1},
Chen Wei\IEEEauthorrefmark{1},
Kexin Lou\IEEEauthorrefmark{1}, 
Dongfeng Gu\IEEEauthorrefmark{1,2}, and
Quanying Liu\IEEEauthorrefmark{1}$^{*}$}
\IEEEauthorblockA{\IEEEauthorrefmark{1}Department of Biomedical Engineering, Southern University of Science and Technology, Shenzhen, China \\
\IEEEauthorrefmark{2}School of Public Health and Emergency Management, Southern University of Science and Technology, Shenzhen, China}
\IEEEauthorblockA{$^*$ Corresponding Author: Quanying Liu \quad Email: liuqy@sustech.edu.cn}}

\maketitle

\begin{abstract}
Electroencephalography (EEG) and Magnetoencephalography (MEG) are pivotal in understanding brain activity but are limited by their poor spatial resolution. 
EEG/MEG source imaging (ESI) infers the high-resolution electric field distribution in the brain based on the low-resolution scalp EEG/MEG observations. However, the ESI problem is ill-posed, and how to bring neuroscience priors into ESI method is the key. Here, we present a novel method which utilizes the Brain Geometric-informed Basis Functions (GBFs) as priors to enhance EEG/MEG source imaging. Through comprehensive experiments on both synthetic data and real task EEG data, we demonstrate the superiority of GBFs over traditional spatial basis functions (e.g., Harmonic and MSP), as well as existing ESI methods (e.g., dSPM, MNE, sLORETA, eLORETA). GBFs provide robust ESI results under different noise levels, and result in biologically interpretable EEG sources. We believe the high-resolution EEG source imaging from GBFs will greatly advance neuroscience research.
\end{abstract}

\begin{IEEEkeywords}
EEG, EEG/MEG source imaging (ESI), Geometry-Informed Basis, Inverse Problem
\end{IEEEkeywords}

\section{Introduction}

EEG and MEG are non-invasive recordings of electrophysiological signals on the scalp. 
Despite their excellent temporal resolution, they suffer from poor spatial resolution of neural activity \cite{baillet2017magnetoencephalography,wei2021edge}. 
To this end, \textit{EEG/MEG source imaging} (ESI) has been developed for inferring the neural sources according to the scalp recordings, which offers a more detailed understanding of neural dynamics and brain function~\cite{liu2018detecting}.

Given the vast number of potential neural sources in the brain and few sensor-level observations, ESI is an ill-posed inverse problem. Additional constraints on brain sources are necessary. Traditional ESI methods (e.g., MNE, sLORETA/eLORETA and dSPM) add regularizations based on neuroscience prior knowledge (e.g., spatial/temporal smoothness or sparsity of brain sources). Rather than using these simple priors, other complex priors have adopted from the brain spatial basis functions (SBFs), including Multivariate Source Pre-localization combined with Data-Driven Parcellization (MSP+DDP)\cite{mattout2005multivariate,chowdhury2013meg}, Gaussian kernels\cite{haufe2011large,wei2021edge}, and Spherical Harmonics\cite{petrov2012harmony}.
However, these SBFs-based ESI methods have their own limitations. Specifically, the MSP+DDP method is data-driven and highly sensitive to noise. The sparse Gaussian kernels cannot cover the whole brain and lack biological interpretability. Harmonic method misses biological information about the brain structure and function.

\begin{figure}[!t]
\centering
\includegraphics[width=8.cm]{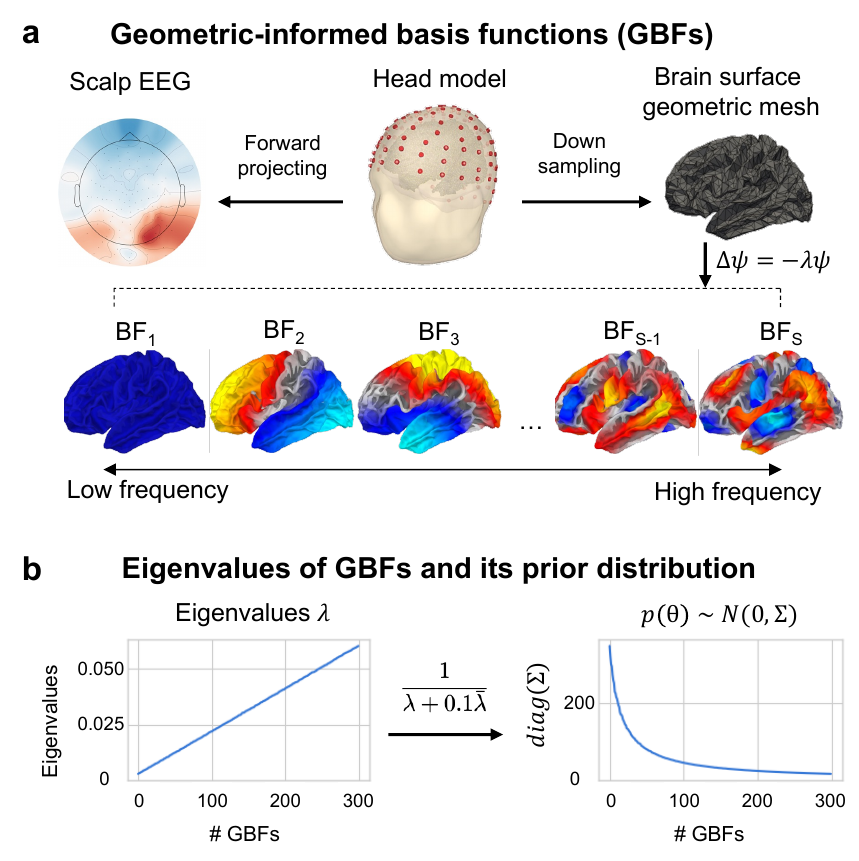}
\caption{\textbf{Geometric-informed basis function for EEG source imaging}. \textbf{a}, Head model (middle) is constructed using individual MRI and BEM method, the scalp EEG (left) can be recovered with forward projecting of EEG source using lead field matrix, and the brain surface geometry mesh (right) is obtained from individual MRI. A series of GBFs (bottom) are calculated by eigendecomposition of the brain geometric. \textbf{b}, The eigenvalues of GBFs (left) and its prior distribution (right). }
\label{fig_GBF}
\end{figure}

Here, we develop the geometric-informed basis functions (GBFs)-based ESI method. GBFs, derived from brain geometry using MRI, are reported to be promising basis functions for fitting fMRI data\cite{pang2023geometric}. These GBFs are task-context independent and are reliable at low signal-to-noise ratios (SNRs). Such properties of GBFs bring unique benefits to ESI. Our study has three main contributions: \\
1) We develop a GBFs-based ESI method that leverages the brain geometric as prior (Fig.~\ref{fig_GBF}). \\
2) We present a framework to generate the EEG source-scalp data from fMRI meta sources (Fig.~\ref{fig_syn}a). \\
3) We demonstrate the superior performance of GBFs using the synthetic data (Fig.~\ref{fig_syn}b \& \ref{fig_noise}) and the real task EEG data (Fig.~\ref{fig_real}).

\section{Methods}
\subsection{EEG/MEG Source Imaging}
ESI is constituted by a forward problem (i.e., head model construction) and an inverse problem (i.e., source localization).

\textit{ESI Forward Problem}:
Mathematically, the forward process from neural sources \( x \) to sensor measurements \( y \) can be formulated as follows:
\begin{equation} \label{eq:forward}
y = Kx + \varepsilon,
\end{equation}
where $x \in\mathbb{R}^{N}$ represents the neural activity in $N$ sources, and $y \in \mathbb{R}^{M} $ denotes the measurements in $M$ sensors, 
and $\varepsilon$ is the measurement noise, $K \in \mathbb{R}^{M \times N}$ is the lead-field matrix, which embodies the head volume conductivity model, describing a linear mapping relationship between sensor signals and source activities ~\cite{vorwerk2018fieldtrip}.

\textit{ESI Inverse Problem}: In Bayesianism, ESI inverse problem is formulated to maximize the posterior probability $P(x | y, K)$, given a prior distribution of brain sources $P(x)$ and the forward model in Eq.~\eqref{eq:forward}, and the scalp measurements $y$. 

\subsection{GBF Algorithm to extract individual GBFs}
Our ESI method employs GBFs~\cite{pang2023geometric}, derived from a cortical surface mesh generated by $Freesurfer$~\cite{fischl2012freesurfer}. We present a GBF algorithm to extract individual GBFs. It takes into consideration the individual cortical geometries to enhance applicability and accuracy. Moreover, it can extract GBFs directly from the source space rather than from T1-weighted (T1W) images. These advantages fulfill critical needs for ESI, allowing EEG researchers to easily employ these extracted GBFs for ESI as they inherently match the source space. 

The computational efficiency of GBF algorithm is improved by down-sampling the mesh (10242 vertices per hemisphere), while keeping adaptability to accommodate various source space sizes.
We use an adapted Laplace–Beltrami Operator (LBO) for the cortex mesh within 3D Euclidean space\cite{seo2011laplace}:
\begin{equation}
    \Delta : = \frac{1}{W}\sum_{i,j} \frac{\partial}{\partial x_i} \left( g^{ij} W \frac{\partial}{\partial x_j} \right)
\end{equation}
where \(x_i, x_j\) are coordinates, \(g^{ij}\) is the inverse of the metric tensor \(g\), and \(W = \sqrt{\text{det}(G)}\) with \(G\) being the matrix representation of \(g\). The metric tensor \(g\) represents the local geometric properties of the cortical surface (e.g., the curvature and shape), describing how distances and angles are measured on the surface at each point. The inverse \(g^{ij}\) is used for transforming vectors between coordinate systems and calculating inner products, ensuring accurate representation of cortical curvature and shape variations.

This LBO operator \(\Delta\) is further used to decompose the cortical mesh into geometric eigenmodes (300 in our study), defined as: $\Delta \psi = -\lambda \psi$, where \(\psi \in\mathbb{R}^{S \times N}\) denotes the geometric eigenmodes(lower panel in Fig.\ref{fig_GBF}.a), where S is the number of the eigenmodes, and \(\lambda \in\mathbb{R}^{S}\) indicates their eigenvalues.  The eigenvalues $\lambda$ are ordered by their spatial frequency, with lower frequencies corresponding to broader spatial patterns, and then transformed into its prior variance for source imaging (See Fig.~\ref{fig_GBF}b).


\begin{figure*}[!h]
\centering
\includegraphics[width=15.5cm]{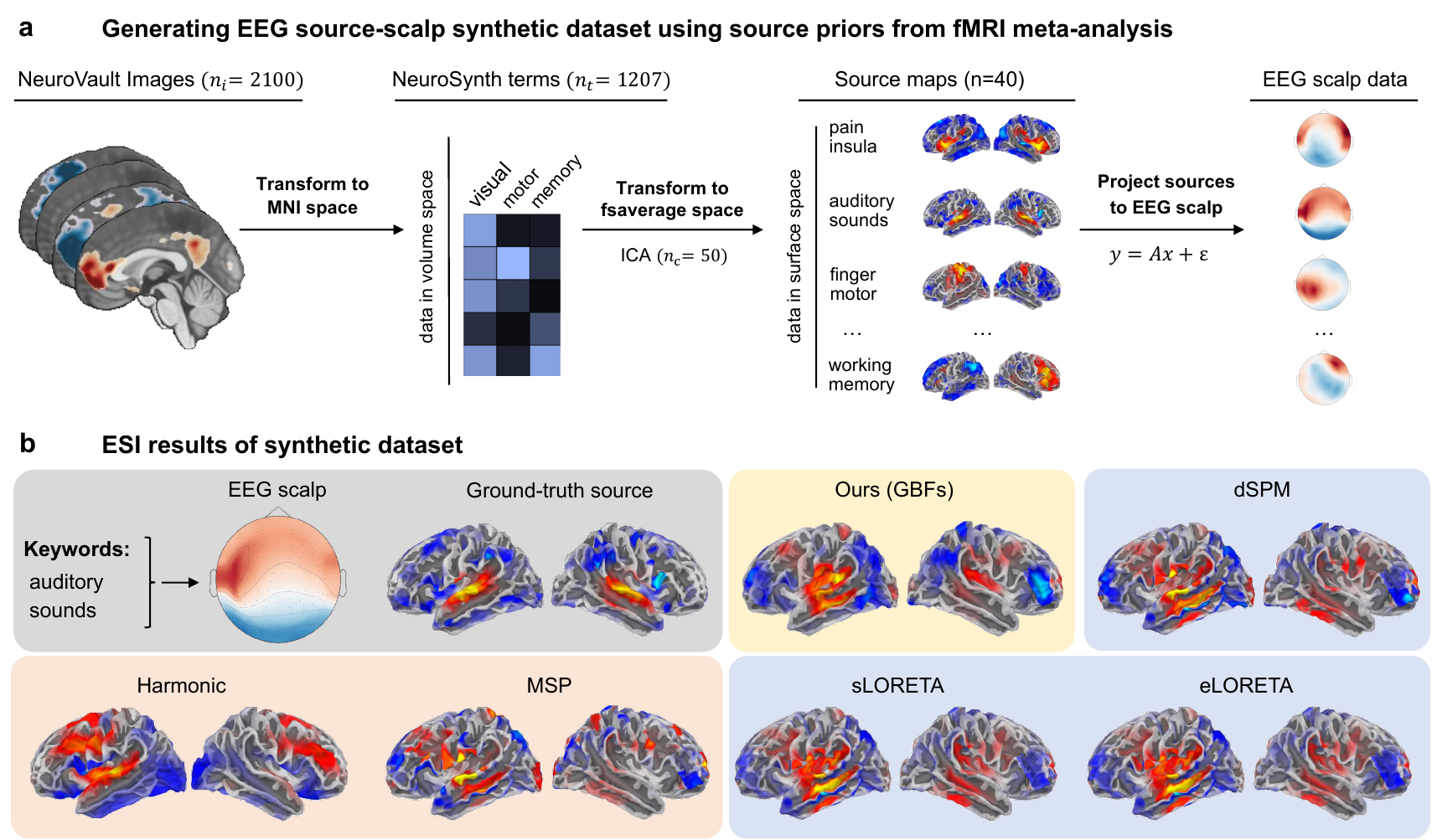}
\caption{\textbf{Generating synthetic dataset and comparisons of ESI results}. \textbf{a}, The EEG source-scalp paired data is generated by fMRI meta-analysis (ICA) from NeuroVault images, transformed from volume space (MNI) to surface space (fsaverage), and finally projected to EEG scalp. \textbf{b}, We exhibit the ground truth (gray), the ESI results from our method (yellow), compared with other basis functions-based methods (red) and traditional ESI methods (blue). }
\label{fig_syn}
\end{figure*}

\subsection{Solving ESI inverse problem with spatial basis functions}
We further employ these spatial basis functions for source imaging, and formulate as
\( x = A \theta \), where \( A \) consists of spatial basis functions and \( \theta \) denotes the corresponding coefficients. For EEG/MEG data \( y \), this relationship is modeled as:
\begin{equation}
    y = K A \theta + \varepsilon.
\end{equation}
We simplify this to \( KA = L \) for further analysis.

Assuming that \( \theta \) follows a Gaussian distribution \( P(\theta) \sim N(0, \Sigma) \), the covariance matrix \( \Sigma \) is defined as a diagonal matrix, with its elements being the reciprocal of eigenvalues of GBFs. As Fig.~\ref{fig_GBF}b, to ensure numerical stability, a regularization term is introduced:
\begin{equation}
    diag(\Sigma) = \frac{1}{\lambda + 0.1 \bar{\lambda}},
\end{equation}
where \( \bar{\lambda} = \frac{1}{L} \sum \lambda \).
The MAP estimation of \( \theta \) is then determined by:
\begin{equation}
    \hat{\theta} = (L^T L + \beta \Sigma^{-1})^{-1} L^T y,
\end{equation}
with \( \beta \) representing the regularization coefficient.

Our approach offers a detailed yet efficient solution for EEG/MEG imaging, enhancing the ESI accuracy by incorporating spatial basis functions. It aligns well with established Minimum Norm methods\cite{he2018electrophysiological} and can be integrated into advanced Bayesian and deep learning frameworks\cite{wei2021edge}.

\section{Experiments}

\subsection{Synthetic source-scalp data}
\subsubsection{Generating source spatial maps based on NeuroVault images}
We present a data generation framework  using the spatial maps from fMRI meta-analysis for creating a  realistic synthetic dataset (Fig.\ref{fig_syn}a). Specifically, we randomly selected 2100 statistical images from NeuroVault~\cite{gorgolewski2015neurovault}, which were transformed into MNI-152 volume space and annotated with 1207 Neurosynth terms \cite{yarkoni2011large}. We construct an imaging-term matrix and run Independent Component Analysis (ICA) to extract 50 spatial maps using $Nilearn$. Then, we manually select 40 spatial maps according to their associated terms, and transform them to fsaverage surface space using $Neuromaps$~\cite{markello2022neuromaps, wu2018accurate}. These maps served as source spatial maps for our synthetic dataset and the corresponding scalp data is obtained by forward projection.

\subsubsection{Noise settings}

\begin{figure}[!t]
\centering
\includegraphics[width=8.8cm]{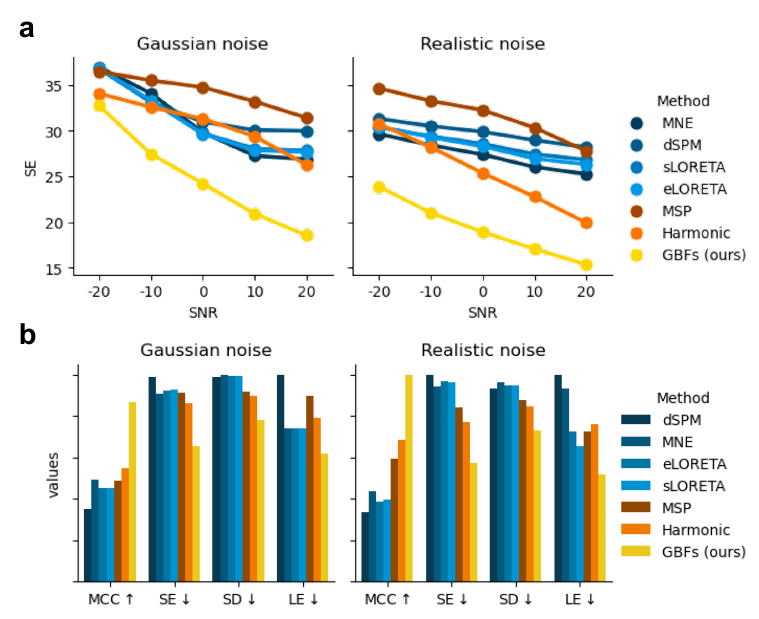}
\caption{\textbf{Comparison of ESI Methods under Various Noise Conditions and SNRs}. \textbf{a}, Comparison of the SE metrics across different SNRs for Gaussian noise (left) and realistic noise (right). Our GBFs-based ESI method consistently outperforms other methods under varying SNR conditions. \textbf{b}, Performance comparison of different ESI methods at SNR=5, in terms of four evaluation metrics. GBFs show superior performance across all metrics compared to other competing methods.}

\label{fig_noise}
\end{figure}

(1) \textit{Gaussian Noise}: 
We assume the sensor noise $n\in\mathbb{R}^{M}$ is channel-independent, and i.i.d. to the normal distribution,
$n   \sim \mathcal{N}(0, I)$.
(2) \textit{Realistic Noise}: 
We generate the realistic noise by using the covariance matrix from the real task EEG data of the VEPCON dataset~\cite{pascucci2022source}. The procedure is encapsulated in the following equation,
$\mathbf{N}_{\text{R}} = \sqrt{\text{diag}(\mathbf{V})} \cdot \mathbf{E} \cdot \mathbf{R}$, where $\mathbf{V}$ and $\mathbf{E}$ are the eigenvalues and eigenvectors of the normalized covariance matrix, and \(\textbf{R}\) is a matrix sampled from standard normal distribution.

\textit{Generating noisy data at different SNRs}:
The generated sensor noise is further scaled to different SNRs, according to
$\text{SNR}_{\text{dB}} = 10 \log_{10} \left( \frac{P_{\text{signal}}}{P_{\text{noise}}} \right)$. Finally,
we mix the clean signal and the noise with SNR ranging from -20 to 20.

\subsection{Competing methods and evaluation metrics}

\subsubsection{ESI methods using other basis functions}
We implemented the ESI methods using other two types of SBFs (i.e., Harmonic and MSP) as competing methods. The Harmonic method conceptualizes each hemisphere as a sphere, projects each source  onto the spherical surface, and then employs Legendre polynomials to solve the spherical harmonic problem, deriving a set of SBFs. In our implementation, we selected polynomials up to the 6th degree, yielding 49 harmonic SBFs. MSP method uses Singular Value Decomposition (SVD) of the whitened lead-field matrix $\mathbf{K}$, and then take the right singular vectors as the MSP spatial basis functions. 

\subsubsection{Evaluation metrics}
For the synthetic data, we have the ground-truth sources. We use Shape Error (SE), Mean Correlation Coefficient (MCC),  Localization Error (LE) and Source Divergence (SD) as evaluation metrics, which have been widely used in previous ESI studies\cite{haufe2011large,wei2021edge}. 





\subsection{The real task EEG data}

We assess the efficacy of our GBFs-based ESI method using two real EEG datasets. (1) The VEPCON dataset includes individual T1 MRI and 128-channel EEG recordings with 2048 sampling rating during a visual task, discriminating faces from scrambled images~\cite{pascucci2022source}. The raw EEG signals were preprocessed with the standard procedure (e.g., downsampled to 250 Hz, detrending, filtering and ICA-based artifact removal). 
(2) The OSE dataset (\url{https://vbmeg.atr.jp/nictitaku209/}) includes the individual T1 MRI and 64-channel EEG recordings during a somatosensory task (i.e., the right median nerve electrical stimulation with eye closed).




\section{Results}
\subsection{ESI results for the synthetic dataset}
Fig.~\ref{fig_syn}b \& Fig.~\ref{fig_noise} show a comprehensive evaluation of our GBFs-based ESI, demonstrating its superiority over all the competing methods, and across various testing conditions. Specifically, Fig.~\ref{fig_syn}b exhibit the estimated sources from 6 ESI methods using synthetic auditory EEG with Gaussian noise at 5 SNR. Our GBFs method uncovers the auditory region highly resemble to the ground truth sources.
In Fig.~\ref{fig_noise}a, we test the robustness of GBFs across a spectrum of SNR levels, ranging from -20 to 20, under Gaussian noise and realistic noise conditions, respectively. The results indicate GBFs reliably outperform all competing ESI methods, in terms of the SE metric. Moreover, Fig.~\ref{fig_noise}b shows the efficacy of GBFs in four evaluation metrics (SE, MCC, SD, and LE) at SNR=5, indicating the superiority of GBF-based ESI in all metrics.

\subsection{ESI results for two real EEG datasets}
\begin{figure}[ht]
\centering
\includegraphics[width=8.2cm]{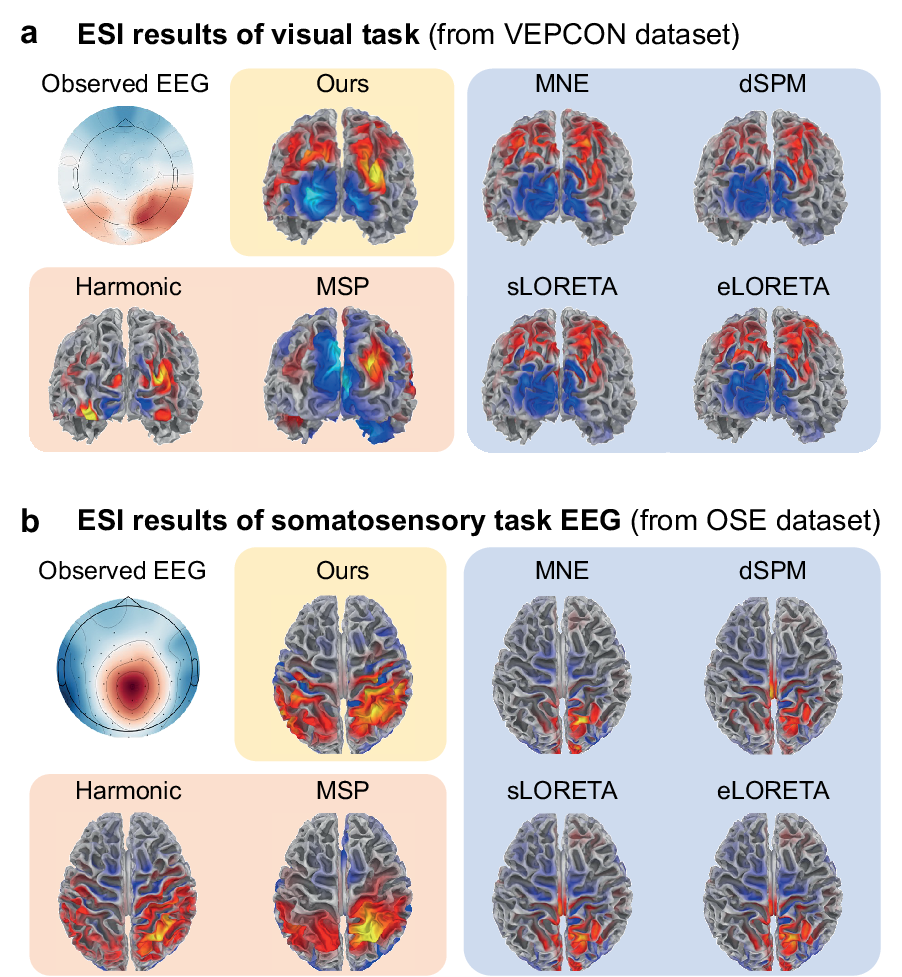}
\caption{\textbf{ESI results for visual task EEG (a) and somatosensory task EEG (b)}. We exhibit the ESI
results from our method (yellow), compared with other basis functions-based method (red), and the traditional EEG source localization
methods (blue).}
\label{fig_real}
\end{figure}

In Fig.\ref{fig_real}, we illustrate the estimated sources from our GBFs-based ESI method, compared with other two SBFs-based ESI (Harmonic and MSP) and four traditional ESI methods (MNE, dSPM, sLORETA and eLORTEA). 
Note that real EEG datasets do not have the ground-truth source images. We expected that the sources of EEG for the visual task and for the somatosensory task would be located in visual areas and somatosensory areas, respectively. The results in Fig.\ref{fig_real} show that the source maps obtained by our GBFs-based ESI method are in good agreement with expectations and consistent with neuroscientific knowledge.

\section{Conclusion}

In summary, we developed the GBFs-based ESI method, using brain geometric priors to advance EEG/MEG source imaging. To validate our method, we presented a framework to synthesize EEG source-scalp data using brain sources from fMRI meta-analysis. Using the synthetic data, we demonstrated the superior performance of GBFs-based ESI over traditional ESI methods and over other spatial basis functions. GBFs can generalize well to the real task EEG data. GBFs-based ESI deals with the spatial resolution limit of EEG/MEG, offering a reliable, biologically interpretable tool for EEG source localization.


\bibliographystyle{IEEEtran}
\bibliography{ref}
\end{document}